\documentclass{elsart}
\usepackage{epsfig}
    \newcommand{\ba}{\begin{eqnarray}}
    \newcommand{\ea}{\end{eqnarray}}
    \newcommand{\be}{\begin{equation}}
    \newcommand{\ee}{\end{equation}}

    \newcommand{\AmS}{{\protect\the\textfont2%
  A\kern-.1667em\lower.5ex\hbox{M}\kern-.125emS}}

\begin{document}
\runauthor{PKU}
\begin{frontmatter}

\title{$I=2$ Pion scattering length with improved actions on anisotropic lattices}

\author[PKU]{Xining Du},
\author[PKU]{Guangwei Meng},
\author[PKU]{Chuan Miao}
\author[PKU]{and Chuan Liu}
\address[PKU]{School of Physics\\
              Peking University\\
              Beijing, 100871, P.~R.~China}

\begin{abstract}
$\pi\pi$ scattering length in the $I=2$ channel is calculated
within quenched approximation using improved gauge and improved
Wilson fermion actions on anisotropic lattices. The results are
extrapolated towards the chiral, infinite volume and continuum
limit. This result improves our previous result on the scattering
length. In the chiral, infinite volume and continuum limit, we
obtain $a^{(2)}_0m_\pi=-0.0467(45)$, which is consistent with the
result from Chiral Perturbation Theory, the experiment and results
from other lattice calculations.
\end{abstract}
\end{frontmatter}


\section{Introduction}

 Low-energy $\pi\pi$  scattering experiment is
 a good testing ground for our understanding
 of the low-energy structure of Quantum Chromodynamics (QCD).
 Chiral Perturbation Theory~\cite{gasser-leutwyler:chiral_oneloop_a},
 Roy equations~\cite{colangelo01:pipi},
 dispersion relations~\cite{zheng02:pipi_dispersion}
 and other theoretical methods have
 been used in the study of low-energy $\pi\pi$ scattering.
 Lattice QCD is a genuine non-perturbative method which
 can handle hadron-hadron scattering at low-energies.
 In recent years, owing to the progress in computing
 facilities and the use of improved actions, there have been
 several studies on pion pion~\cite{jlqcd99:scat,%
 chuan02:pipiI2,JLQCD02:pipi_length,juge03:pipi_length,%
 ishizuka03:pipi_length,CPPACS03:pipi_phase},
 pion nucleon~\cite{fukugita95:scat},
 pion kaon~\cite{chuan04:Kpi} and
 kaon nucleon~\cite{chuan04:KN} scattering in quenched lattice QCD.
 Very recently, CP-PACS collaboration even calculated
 pion pion scattering phase shift in the $I=2$ channel using
 their unquenched configurations~\cite{CPPACS03:pipi_phase_unquench,CPPACS04:pipi_phase_unquench}.
 In this letter, we report our recent quenched lattice
 result on the $\pi\pi$ scattering length in the $I=2$ channel.
 The result in this letter is an improvement over our
 previous result on the scattering length~\cite{chuan02:pipiI2}.

 The lattice gauge action used in our study is
 the tadpole improved action on anisotropic
 lattices~\cite{colin97,colin99}.
 Using this gauge action together with improved Wilson fermion actions,
 glueball and hadron spectra have been studied
 within quenched approximation \cite{colin97,colin99,%
 chuan01:gluea,chuan01:glueb,chuan01:canton1,chuan01:canton2,chuan01:india}.
 Configurations generated from the
 gauge action have also been utilized to calculate the
 the scattering length for
 $\pi\pi$ scattering in the $I=2$ channel~\cite{chuan02:pipiI2},
 $KN$ scattering in the $I=1$ channel~\cite{chuan04:KN} and
 the $K\pi$ scattering in the $I=3/2$ channel~\cite{chuan04:Kpi}.
 In this letter, we update our result on the
 $\pi\pi$ scattering length in the $I=2$ channel.
 The number of gauge field configurations has
 been increased. A more general chiral extrapolation method is used.
 Otherwise, the basic procedure of the calculation is
 similar to that adopted in Ref.~\cite{chuan02:pipiI2}.

 \section{Numerical calculation of the scattering length}
 \label{sec:theory}

 Configurations used in this calculation are generated using the pure
 gauge action for $6^340$, $8^340$ and $10^350$ lattices with the gauge
 coupling $\beta=1.9$, $2.2$, $2.4$, $2.6$ and $3.0$. The spatial lattice
 spacing $a_s$ is roughly between $0.1$fm and $0.4$fm while
 the spatial physical size of the lattice ranges from
 $0.7$fm to $4.0$fm. For each set of parameters, several hundred
 (typically $512$) decorrelated gauge field configurations
 are used to measure the fermionic quantities.
 Statistical errors are analyzed using the jack-knife method.
 Single pion, and rho  mass values
 are obtained from the plateau of their corresponding effective
 mass plots with the fitting interval being automatically
 chosen by minimal $\chi^2$ per degree of freedom.

 In order to calculate the elastic scattering lengths
 for hadron-hadron scattering on the lattice,
 or the scattering phase shifts in general, one uses L\"uscher's
 formula which relates the exact energy level of two hadron states
 in a finite box to the elastic scattering phase shift in the continuum.
 For $\pi\pi$ scattering at zero relative three momentum,
 this formula amounts to a relation
 between the {\em exact} energy $E^{(I)}_{\pi\pi}$ of the
 two pion system with vanishing relative momentum
 in a finite box of size $L$ with isospin $I$, and
 the corresponding scattering length
 $a^{(I)}_0$ in the continuum.
 This formula reads \cite{luscher91:finitea}:
 \be
 \label{eq:luescher}
 E^{(I)}_{\pi\pi}-2m_\pi=-\frac{4\pi a^{(I)}_0}{m_\pi L^3}
 \left[1+c_1\left(\frac{a^{(I)}_0}{L}\right)
 +c_2\left(\frac{a^{(I)}_0}{L}\right)^2
 \right]+O(L^{-6}) \;\;,
 \ee
 where $c_1=-2.837297$, $c_2=6.375183$ are numerical constants
 and $m_\pi$ is the  mass of the pion.
 In this letter, we focus on the $\pi\pi$ scattering length $a^{(2)}_0$
 in the isospin $I=2$ channel.

 To measure the hadron mass values $m_\pi$, $m_\rho$ and
 to extract the energy shift $\delta E^{(2)}_{\pi\pi}$,
 one constructs the correlation
 functions from the corresponding one meson and two meson operators
 in the appropriate symmetry channel.
 In this letter, we used the same operators as those
 in Ref.~\cite{chuan02:pipiI2}.
 Numerically, it is more advantageous to construct
 the ratio of the correlation functions:
 \be
 {\mathcal R}^{I=2}(t) = C^{I=2}_{\pi\pi}(t) / (C_\pi(t)C_\pi(t))
 \;\;,
 \ee
 where $C^{I=2}_{\pi\pi}(t)$ is the two-pion correlation
 function in the $I=2$ channel and $C_\pi(t)$ is the
 one-pion correlation function.
 One then uses the  fitting function:
 \be
 \label{eq:linear_fit}
 {\mathcal R}^{I=2}(t) \propto
 e^{-\delta E^{(2)}_{\pi\pi}t}\sim 1-\delta E^{(2)}_{\pi\pi}t
 \;\;,
 \ee
 to determine the energy shift $\delta E^{(2)}_{\pi\pi}$.
 Usually, the linear form is sufficient. However, for small
 lattices, the exponential form should be used. Similar
 situation was also observed by the JLQCD and CP-PACS
 collaboration in their
 calculations~\cite{JLQCD02:pipi_length,CPPACS03:pipi_phase}.

 Two pion correlation function, or equivalently, the ratio
 ${\mathcal R}^{I=2}(t)$ is constructed from quark propagators,
 which are obtained using the Multi-mass Minimal
 Residue algorithm with wall sources.
 Periodic boundary condition is applied to all three
 spatial directions while in the temporal direction, Dirichlet
 boundary condition is utilized.

 After obtaining the energy shifts $\delta E^{(2)}_{\pi\pi}$
 from the two-pion correlation functions,
 the values of $\delta E^{(2)}_{\pi\pi}$are
 substituted into L\"uscher's formula to solve
 for the scattering length $a^{(2)}_0$ for all parameter sets
 that have been simulated.
 From these results, attempts are made to perform
 an extrapolation towards the chiral, infinite volume
 and continuum limit.

 \section{Extrapolations of the scattering length}
 \label{sec:extrap}

 The chiral extrapolations of physical quantities are performed
 as discussed in Ref.~\cite{chuan02:pipiI2}.
 The only difference is that, for all physical quantities,
 since we now have more bare quark
 mass values available,
 we tried to use a quadratic extrapolation
 in the bare quark mass rather than a linear extrapolation
 as used in Ref.~\cite{chuan02:pipiI2}.
 For each set of parameters, we have calculated
 fermionic quantities with $8$ different values of $\kappa$.
 The largest value of $\kappa$, which corresponds to lowest
 pion mass for each simulation point is chosen such that the solution
 of the quark propagator can be obtained within a reasonable number
 of Minimal Residue iterations (typically $600$ to $800$).
 The lowest pion to rho mass ratio for our simulation
 points are between $0.66$ and $0.74$.
 For the scattering length $a^{(2)}_0$,
 we use the quantity~\cite{chuan02:pipiI2}
 $F=a^{(2)}_0m^2_\rho/m_{\pi}$, which in the chiral
 limit is given by the current algebra
 value:~\cite{weinberg66:pipi_CA}
 \be
 \label{eq:chiral}
 F\equiv {a^{(2)}_0m^2_\rho \over m_{\pi}}
 =-{1 \over 16 \pi} {m^2_\rho \over f^2_\pi}
 \sim  -1.364\;\;.
 \ee
 Here the final numerical value is obtained
 by substituting in the experimental values for
 $m_\rho\sim 770$MeV and $f_\pi\sim 93$MeV.
 Chiral perturbation theory with loops yields almost
 the same numerical value for factor $F$~\cite{colangelo01:pipi}
 in the $I=2$ channel.
 Note that the factor $F$ can be calculated on the lattice with good
 precision {\em without} the lattice calculation of meson
 decay constants.
 We have adopted a quadratic functional form in chiral
 extrapolation for all physical quantities and
 the fitting range of the extrapolation is self-adjusted by the
 program to yield a minimal $\chi^2$ per degree of freedom.

 \begin{figure}[thb]
 \begin{center}
 \includegraphics[height=10.0cm,angle=0]{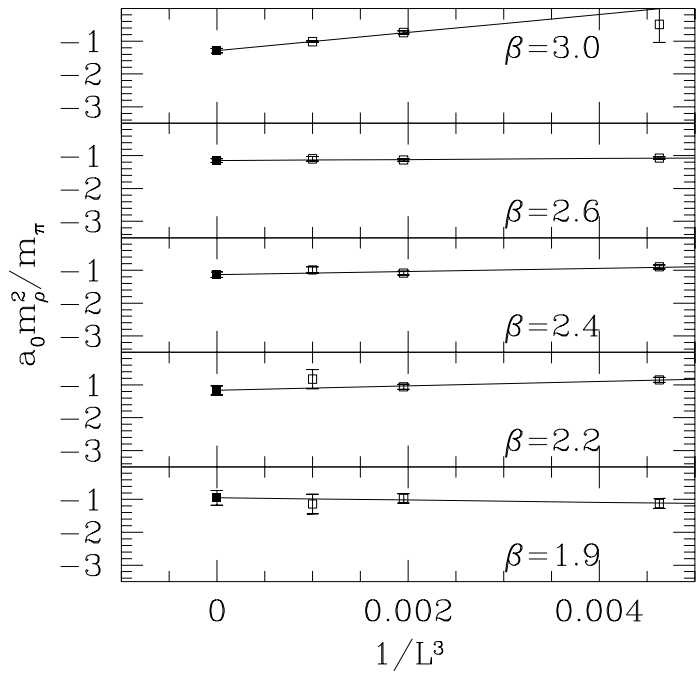}
 \end{center}
 \caption{Infinite volume extrapolation for the quantity
 $F=a^{(2)}_0m^2_\rho/m_{\pi}$ obtained from our simulation results
 at $\beta=3.0$, $2.6$, $2.4$, $2.2$ and $1.9$.
 The straight line represents the linear extrapolation in $1/L^3$.
 The extrapolated results are shown as solid square with
 the corresponding error at $L^{-3}=0$.
 \label{fig:volume_extrapolation}}
 \end{figure}
 The values of $F$ after the chiral extrapolations are
 used to extrapolated towards the
 infinite volume limit by a function linear in $1/L^3$,
 as suggested by L\"uscher's formula.
 This extrapolation is shown in Fig~\ref{fig:volume_extrapolation}
 for all five values of $\beta$.
 The extrapolated results thus obtained for the factor $F$ are
 then used for the continuum limit extrapolation.

 \begin{figure}[thb]
 \begin{center}
  \vspace{-50mm}
 \includegraphics[height=10.0cm,angle=0]{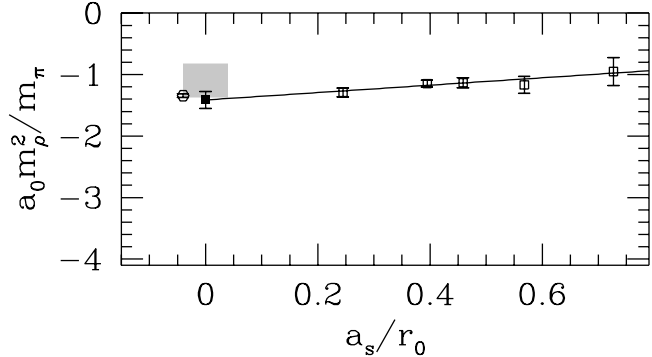}
 \end{center}
 \caption{Continuum extrapolation for the quantity
 $F=a^{(2)}_0m^2_\rho/m_{\pi}$ obtained from our simulation results
 at $\beta=3.0$, $2.6$, $2.4$, $2.2$ and $1.9$.
 The straight line represents the linear extrapolation in $a_s/r_0$.
 The extrapolated result is also shown as solid square with
 the corresponding error. Also shown near $a_s=0$ are the
 (two-loop) result from chiral perturbation theory~\cite{colangelo01:pipi}
 (open hexagon) and the experimental result (shaded region).
 \label{fig:continuum_extrapolation}}
 \end{figure}
 Finally, an continuum limit extrapolation is performed
 to eliminate the finite lattice
 spacing errors. Since we have used the tadpole
 improved clover Wilson action, all physical quantities
 differ from their continuum counterparts by
 terms that are proportional to the spatial lattice spacing $a_s$.
 The physical value of $a_s$ in terms of the hadronic scale $r_0$
 for each value of $\beta$ can be
 found from Ref. \cite{colin99,chuan01:india}.
 The result of the continuum extrapolation is shown in
 Fig.~\ref{fig:continuum_extrapolation} where
 the results from the chiral and infinite volume
 extrapolation discussed above are indicated as
 data points (open squares with error bars)
 in the plot for all $5$ values of
 $\beta$ that have been simulated.
 The straight line shows
 the extrapolation towards the $a_s=0$ limit and the
 final extrapolated result is also shown as a solid square.
 For comparison, the corresponding result from
 two-loop chiral perturbation theory~\cite{colangelo01:pipi}
 is also shown near $a_s=0$ as the open hexagon.
 The current algebra result and
 the one-loop chiral perturbation result are quite close
 to the two-loop chiral result numerically. To avoid crowdedness,
 they are not shown in the figure.
 The experimental result from E865 Collaboration~\cite{E86500:pipi}
 is shown as the shaded region near $a_s=0$.
 The height of the shaded region designates the error for
 the experimental result.
 It is evident that these results are compatible within error bars.

 To summarize, we obtain from the continuum extrapolation the following
 result: $F=a^{(2)}_0m^2_\rho/m_{\pi}=-1.41(14)$.
 If we substitute in the physical meson mass values, we obtain the
 $\pi\pi$ scattering length in the $I=2$ channel as:
 $a^{(2)}_0m_\pi=-0.0467(45)$, which is to be compared
 with the current algebra value of $-0.046$~\cite{weinberg66:pipi_CA},
 one-loop chiral perturbation theory result
 of $-0.042$~\cite{gasser-leutwyler:chiral_oneloop_a}
 and the two-loop result of $-0.0444(10)$~\cite{colangelo01:pipi}.
 Our result is also consistent with the experimental result
 from E865 Collaboration which is:
 $a^{(2)}_0m_\pi= -0.036(9)$~\cite{E86500:pipi}.
 The result reported in this letter is also consistent
 with our previous result with the error reduced by a factor of
 two or so due to higher statistics and lattices with finer lattice spacing.
 This result also agrees with other lattice
 studies reported
 recently~\cite{JLQCD02:pipi_length,CPPACS03:pipi_phase}.
 Their final result for $a_0m_\pi$ is:
 $a^{(2)}_0m_\pi= -0.0410(69)$, which is also compatible
 with our result within error.

 \section{Conclusions}
 \label{sec:conclude}

 In this letter, we report our lattice result on pion-pion scattering
 lengths in isospin $I=2$ channel, obtained using quenched
 lattice QCD.  Simulations are performed on lattices
 with various sizes, ranging from $0.7$fm to about
 $4$fm and with five different values of lattice spacing
 with several hundred gauge field configurations.
 Quark propagators are measured with $8$ different
 valence quark mass values. These enable us to
 explore the chiral limit, the finite volume and the finite
 lattice spacing errors in a systematic fashion.
 The lattice result for the scattering length is
 extrapolated towards the chiral, infinite volume
 and continuum limit where a result consistent with
 the experiment, Chiral Perturbation Theory and previous
 lattice results is found.

 \section*{Acknowledgments}

 This work is supported by the National Natural
 Science Foundation of China (NSFC) under grant
 No. 90103006, No. 10235040 and by the Trans-century fund from
 the Chinese Ministry of Education. C. Liu would like
 to thank Prof.~H.~Q.~Zheng and for helpful discussions.


\end{document}